\documentstyle[pre,aps,psfig]{revtex}

\newcommand \beq{\begin{eqnarray}}
\newcommand \eeq{\end{eqnarray}}
\newcommand{\set}[2]{\newcommand{#1}{#2}}
\set{\pa}{\partial \over \partial\, }
\set{\leftvector}{\stackrel{\leftarrow}{\partial }}
\set{\rightvector}{\stackrel{\rightarrow}{\partial }}

\begin{document}
\twocolumn[\hsize\textwidth\columnwidth\hsize
           \csname @twocolumnfalse\endcsname

\title{Consequences of coarse grained Vlasov equations}

\author{Klaus Morawetz\dag, Rainer Walke\ddag}

\address{\dag\ 
Max-Planck-Institute for the Physics of Complex Systems, 
Noethnitzer Str. 38, 01187 Dresden, Germany}
\address{\ddag\  Max-Planck-Institute for
    demographic research, Rostock, Germany
}

\maketitle

\begin{abstract}
The Vlasov equation is analyzed for coarse grained distributions
resembling a finite width of test-particles as used in numerical
implementations. It is shown that this coarse grained distribution obeys a
kinetic equation similar to the Vlasov equation, but with additional terms.
These terms give rise to entropy production
indicating dissipative features due to a nonlinear mode coupling 
The interchange of coarse graining and dynamical
evolution is discussed with the help of an exactly solvable model for
the selfconsistent Vlasov equation and practical consequences are worked out.
By calculating analytically the stationary solution of a general 
Vlasov equation we
can show that a sum of modified Boltzmann-like distributions is approached dependent on the initial distribution. This
behavior is independent of degeneracy and only controlled by the
width of
test-particles.  The condition for approaching a stationary solution
is derived and it is found that the coarse graining energy given by the
momentum width of test particles should be smaller than a quarter of
the kinetic energy.  
Observable consequences of this coarse graining are:
(i) spatial correlations in observables, (ii) too large radii of
clusters or nuclei in self-consistent Thomas-Fermi treatments, (iii) a

structure term in the response
function resembling vertex correction correlations or internal
structure effects and (iv) a modified centroid energy and higher
damping width of collective modes. 
\end{abstract}




\vskip2pc]

\section{Introduction}

The Boltzmann kinetic equation is a successfully used dynamical model to
describe heavy ion
collisions and multifragmentation up to intermediate energies
\cite{BCR92,P89,HM87,BKG84,BD88,D84,GRSVR87,Ba92,AB85,KWB92,MLSCN98}.
The scattering is added to the Vlasov equation either by the relaxation time 
\cite{KN84,D84a,K85,AK93,KB89} or by scattering 
integrals of the Boltzmann-type
\cite{BKG84,GRSVR87,KB89,N28,UU33,GBD87,KJS85,SG86,CLG87} or
their nonlocal extensions \cite{SLM96,MLSK98,MLNCCT01}
in the same way as it was done in the 
Landau-Vlasov equation for hot plasmas. All these simulation have in
common that the motion of particles between collisions are covered by
Hamilton equations or the Vlasov kinetic equation with a selfconsistent potential. 

Even within
collisionless Vlasov codes the onset of fragmentation is described
\cite{BKN76,W82,BBS84}. The use of selfconsistent Vlasov equation is
not restricted to nuclear collisions but has been applied successfully
for collisions of ions with metal clusters \cite{GG96,PG99,CRSU00}.
Despite
the fact that the Vlasov equation is a reversible kinetic equation and the
initial configuration should be retained after a long enough calculation,
fragmentation and energy dissipation is observed. Obviously
this is due to the fact that the Poincar{\' e} time is much larger than the time
where the phase space is filled by various trajectories. Therefore one can
consider superficially this spreading as an irreversible process in the sense of entropy production.
Nevertheless, the underlying dynamics is reversible.

The numerical implementation of
Vlasov codes demands a certain coarse graining of the space and momentum
coordinates. This numerical uncertainty is quite sufficient to generate
genuine dissipation and entropy production. This fact has been
investigated in \cite{JB96} and will be considered in detail within
this paper. It has been argued that the errors due to numerically
coarse graining accumulate diffusively. We will demonstrate that these
errors will lead to a unique equilibration of Vlasov equation. While
the diffusion itself is less effected, the coarse graining leads to a different
dynamical evolution and has consequences for the extraction of damping
rates from Vlasov-type simulations.

Theoretically one can derive kinetic equations by
phase-space averaging of the Liouville equation which itself is exactly of Vlasov type \cite{K75}.
The resulting collision integrals represent two different approximations of the
nonequilibrium dynamics: (i) The truncation of coupling to higher order
correlations (hierarchy) and (ii) The smoothing procedure which translates
the fluctuating stochastic equation into a kinetic equation for the smoothed
distribution function. This latter procedure is sometimes also called coarse
graining.
For the discussion of appropriate collision
integrals also in the quantum case see \cite{LSM97}. Here we will focus on a detailed analysis
of consequences of coarse
graining. 

The idea used here dates back to the work of Gibbs and Ehrenfest \cite{G5,EE7}.
They suggested to coarse grain the entropy definition by a more rough distribution function
\beq
f(p,r,t)={1 \over \Delta(p,r) } \int\limits_{\Delta(p,r)} f_{\delta}(p',r',t) dp' dr'.\label{fd}
\eeq
The physical meaning consists in the fact that any observable is a mean value of an averaging about a certain area in phase space. It was shown that the entropy with this coarse grained distribution increases \cite{S76}. This means that in a closed system the entropy can rise if we average the observation about small phase space cells.

The interpretation is that other phase space points can enter and leave the cell which is not compensated. A phase space mixing occurs \cite{S76} since the two limits cannot be interchanged, i.e. the thermodynamic limit and the limit of vanishing phase space cell. The coarse graining of Ehrenfest can be observed if the thermodynamical limit is carried out first and the limit of small phase space cells afterwards. It solves therefore not the problem of entropy production, but gives an interesting aspect to entropy production by coarse grained observations \cite{S76}.

In this paper we like to investigate three questions:
\begin{itemize}
\item
Which kinetic equation is really solved numerically if the Vlasov equation is implemented in numerics?
\item
What are the properties of this kinetic equation, especially which kind of dissipative features appear ?
\item
What are the consequences to practical applications, e.g. damping of giant resonances and binding energies?
\end{itemize}

The outline of the paper is as follows. Next we derive the kinetic
equation which is obeyed by the coarse grained distribution
function. Then in chapter III we discuss the entropy production. We
demonstrate with the help of two exactly solvable models that this entropy production is due to mixing, i.e. a mode coupling and not simply by spreading of Gaussians. The solution of the stationary Vlasov equation is then presented in chapter IV. We will find that the stationary solution can be represented as an infinite sum of modified Boltzmann distributions. This expansion shows the unique character of time evolution which is only determined  by the initial distribution. In chapter V we discuss consequences of this result: (i) The thermodynamics becomes modified by spatial correlations. The selfconsistency leads to a modified Thomas- Fermi equation lowering the binding energy, (ii) the structure factor shows a substructure similar as obtained from vertex corrections and (iii) the damping width
of collective resonances is shown to be larger by coarse graining. While the centroid energy is smaller by momentum coarse graining it increases by spatial coarse graining.

\section{Coarse grained Vlasov equation}

The origin of the coarse graining may be the numerical implementation or the use of averaged distribution functions instead of the fluctuating one. To illustrate the method we examine the quasi-classical Vlasov equation and show which equation is really solved if one is forced, by numerical reasons, to use coarse graining. It will become clear shortly that instead of Vlasov equation a modified kinetic equation is solved when coarse graining is present. The quantum mechanical or TDHF equation can be treated in analogy. We start from the Vlasov equation
\begin{equation}
  {\pa t} f_{\delta}(prt) +{p \over m} {\pa r} f_{\delta}(prt) -{\pa r} V_{\delta}(r,t)
{\pa p}  f_{\delta}(prt)=0
\label{vlas}
\end{equation}
which solution can be formally represented as an infinite sum of exact test-particles
\begin{equation}
  f_{\delta}(prt)=\sum\limits_{i=1}^{\infty} \delta(r-R_i(t))\,\delta
  (p-P_i(t))\label{sum}
\end{equation}
where the test-particle positions and momenta evolve corresponding to
the Hamilton equations $ {\dot R_i(t)}=P_i/m$ and $ {\dot
  P_i(t)}=-\partial_R V(R_i(t))$. In the following we understand
$p,r,P$ and $R$ as vectors and suppress their explicit notation. 

For the mean-field term $V_{\delta}$ we assume a Hartree approximation given by a convolution of the density with the two-particle interaction $V_0$
\begin{equation}
  V_{\delta}(r,t)= \int dr' V_0(r-r') \int {dp' \over (2\pi \hbar)^3}
    f_{\delta}(p'r't).\label{vd}
\end{equation}
In practice all numerical calculations use two assumptions : (i) The infinite number of test-particles is truncated by a finite value. (ii) The used test particle has a finite width due to numerical errors and/or smoothing demand of the procedure. While in \cite{RS95a,RS95b,LSR95} was discussed that the approximation (i) leads to a Boltzmann-like collision integral, the approximation (ii) will deserve further investigations. Especially, we will show that the finite width of test-particles leads to a coarse graining and a dissipation forcing the system to a Boltzmann- like distribution. This is even valid with infinite numbers of test-particles. Therefore we consider the effect of coarse graining as the most determining one for one-body dissipation.

The finite width of test particles can be reproduced most conveniently by a convolution of the exact solution (\ref{sum}) with a Gaussian $g_a(x)=(2 \pi \sigma_a^2)^{-1.5} {\rm exp} (-x^2/(2 \sigma_a^2))$ resulting in the coarse grained distribution function $f$
\begin{eqnarray}
  f(prt)&=&\{f_{\delta}\}_g\nonumber\\
&=&\int {d p' \over (2 \pi \hbar)^3} d r'
  g_r(r-r') g_p(p-p') f_{\delta}(p'r't)\nonumber\\
&=&\sum\limits_{i=1}^{\infty} g_r(r-R_i(t))g_p(p-P_i(t)).\label{f}
\end{eqnarray}

To answer the question which kind of kinetic equation describes this smoothed distribution function,
the kinetic equation for (\ref{f}) is derived from (\ref{vlas}) by
convolution with a Gaussian. The equation for general coarse grained
mean-fields has been already derived in \cite{LSR95}. In order to make
the physical content more explicit we calculate the different terms
explicitly and present the coarse grained kinetic equation. The free drift term ${p \over m} \partial_r f_{\delta}$ takes the form after convolution
\begin{equation}
  \{{p \over m} \partial_r f_{\delta}\}_g={p \over m} {\pa r} f(prt)
  +{\sigma_p^2 \over m} {\pa r}{\pa p} f(prt)\label{pm}
\end{equation}
which is established by partial integration. We see that the free
streaming is modified by an additional resistive term which will give
dissipative features.

The mean-field term takes the form
\begin{eqnarray}
\{{\pa r} V_{\delta}(r){\pa p} f_{\delta}(prt)\}_g&=&{\pa p}\{{\pa r}
V_{\delta}(r) f_{\delta}(prt)\}_g\nonumber\\
=&&{\pa p}\{\{{\pa r} V_{\delta}(r) f_{\delta}(prt)\}_{g_r}\}_{g_p}.
\end{eqnarray}
The space convolution with $g_r$ is performed using the relation \cite{T89}
\begin{equation}
  \{AB\}_g=\{A\}_g \rm{exp}(\sigma_r^2 {\leftvector}_r {\rightvector}_r)\{B\}_g\label{rel}
\end{equation}
with the result
\begin{equation}
  {\pa p}\{{\pa r} V(r,t) \exp (\sigma_r^2 {\leftvector}_r
    {\rightvector}_r) \{f_{\delta}(prt)\}_{g_r}\}_{g_p}.\label{vp}
\end{equation}
$V(r,t)$ is the mean-field
calculated with the space and momentum coarse grained distribution $f$
instead of $f_{\delta}$, via (\ref{vd}) which can be seen as 
\begin{eqnarray}
V(r,t)&=& \{V_{\delta}(r,t)\}_{g_r}=\int dr'' V_0(r'')  \int dr' g_r(r-r''-r')\nonumber\\
&\times& \int
  {dp' \over (2\pi \hbar)^3} f_{\delta}(p',r',t)\nonumber\\ &=&\int dr''
  V_0(r'') \int dr' g_r(r-r''-r') \nonumber\\
&\times&\int {dp' \over (2\pi \hbar)^3} \int
  {dp \over (2\pi \hbar)^3} g_p(p-p') f_{\delta}(p',r',t)\nonumber\\
  &=&\int dr'' V_0(r'') \int {dp \over (2\pi \hbar)^3} f(p,r-r'',t)\label{vr}
\end{eqnarray}
where the last equality shows the invariance of particle density due to
coarse graining.
The momentum convolution with the Gaussian is then performed in (\ref{vp})
to yield the momentum and space coarse distribution function $f$.
We like to point out that the test-particle method would lead to a
further folding of the mean-field potential if it is read off from a finite grid \cite{RS95a}.

The coarse grained Vlasov equation reads now
\begin{eqnarray}
  {\pa t} f(prt)&+&{p \over m} {\pa r} f(prt)+{\sigma_p^2 \over m} {\pa
    r}{\pa p} f(prt)\nonumber\\
&-&{\pa r} V(r,t) \exp (\sigma_r^2 {\leftvector
    }_r {\rightvector}_r) {\pa p} f(prt)=0.\label{kinetic}
  \end{eqnarray}
This equation is the main result of this chapter and represents the Vlasov
equation for the coarse grained distributions and should be compared
with the Husimi representation of \cite{LSR95}. While the distribution
function is exactly the Husimi representation of the Wigner function
the corresponding coarse grained kinetic equation was not given before.

Equation (\ref{kinetic}) represents the kinetic equation which is really solved numerically if the Vlasov equation is attempted to be solved. The coarse graining leads
to two additional contributions besides the original Vlasov equation. We will
see that this causes just the dissipative like features. While we will
continue now and investigate (\ref{kinetic}) more closely, in appendix
\ref{app} the question is addressed how the underlying dynamics is
modified by coarse graining. We find that in principle the coarse
grained equation can be mapped to the original Vlasov equation if one
defines new testparticles with modified dynamical equations by a
modified potential. The relation between the original and the modified
potential appears just as the inverse folded mean field potential which is the opposite relation than presented in \cite{RS95a,RS95b,LSR95}.

\section{Entropy production}

To analyze the dissipative feature we rewrite (\ref{kinetic}) into
\begin{equation}
  {\pa t} f(prt) +{p \over m} {\pa r} f(prt) -{\pa r} V(r,t)
{\pa p}  f(prt)=I_{\rm diss}
\label{vlas1}
\end{equation}
with the one-body (collision integral like) term
\begin{eqnarray}
  I_{\rm diss}&=&-{\sigma_p^2 \over m} {\pa r}{\pa p} f(prt)\nonumber\\
&+&{\pa r}
  V(r,t) (\exp (\sigma_r^2 {\leftvector
    }_r {\rightvector}_r)-1) {\pa p} f(prt)\nonumber\\
    &\approx& (\sigma_r^2{\partial^2 \over \partial r^2} V(r,t)-{\sigma_p^2 \over
      m}) {\pa r}{\pa p} f(prt) + o(\sigma^4).\nonumber\\\label{coll}
\end{eqnarray}

In order to demonstrate the entropy production explicitly we use the linearized form of (\ref{coll}) and built up the balance equation for entropy $S=f {\rm ln} f$ by multiplying (\ref{vlas1}) with ${\rm ln} f$ and integrating over $p$. The entropy balance reads then
\begin{equation}
  {\dot S(r,t)}+{\pa r} \int {dp \over (2\pi \hbar)^3} {p \over m} f
  {\rm ln}f = \Phi(r,t) \int {dp \over (2\pi \hbar)^3} {\partial_r f
    \partial_p f \over f}\label{entropy}
\end{equation}
with $\Phi(r,t)=-\sigma_r^2{\partial^2 \over \partial r^2} V(r,t)+{\sigma_p^2 \over m}$.
The entropy density obeys therefore a conservation law with a source term on
the right hand side of (\ref{entropy}). Especially one sees that the total
entropy change is  equal to
\begin{equation}
  {\dot S(t)}= \int dr {\dot S}(r,t)=\int dr \Phi(r,t) \int {dp \over (2\pi \hbar)^3} {\partial_r f
    \partial_p f \over f}.\label{entropy1}
\end{equation}
This expression allows already to learn some interesting properties of such distribution functions which lead to an entropy increase.
For a distribution function symmetric either in $p$ or $r$, no entropy production occurs since $\partial f$ would be asymmetric. We get an entropy production only for explicit space dependent distributions due to the $\partial_r f$ factor.
Near equilibrium where the assumed nonequilibrium
distribution functions are falling in space and
momentum, the second derivative of the mean-field is negative (due to
finite systems) and therefore $\Phi>0$. Consequently we obtain an increase of entropy on average due to the
spatial coarse graining. This means that the finite width of test-particles
induces an entropy increase similar like irreversibility. 

The reason for the entropy increase is a nonlinear mode coupling which
can be considered as a general feature of coarse graining.
Therefore we rewrite the collision integral in the first line of
(\ref{coll})  by Fourier transformation into another
form
\begin{eqnarray}
  I_{\rm diss}&=&\int dr' {\pa p} f(r'pt) \left [ \delta(r-r')
  ({\sigma_p^2 \over m} \leftvector_r-{\pa r} V(r,t) )\right.\nonumber\\
&+&\left.
 \int dr'' {\pa r''} V(r'') \int {dk_1 dk_2
    \over (2\pi \hbar)^6} \right .\nonumber\\
&\times&\left .{\rm exp}(-i(k_1+k_2) r-\sigma_r^2 k_1 k_2+i
  k_2 r' +i k_1 r'')\right].\nonumber\\
&&
\label{dissp}
\end{eqnarray}
We see that due to the spatial coarse graining a nonlinear mode
coupling occurs. This is represented by the product $k_1 k_2$ between
the modes in the distribution function and the mean field. This latter
effect is the reason for the production of entropy and is connected
with the spatial coarse graining $\sigma_r$.\footnote{Please remark
  that $f$ and $V$ are coarse grained values itself which Fourier-transforms
  contain an exponential $k_1^2 \sigma_r^2/2$ and $k_2^2
  \sigma_r^2/2$ respectively. Therefore the total exponent
  $(k_1+k_2)^2\sigma_r^2/2$ appears and the expression is convergent
  while the bare $k_1 k_2$ product above may lead to
  the impression of non-convergence.}

To understand the physical origin of this entropy production we choose two simple models for illustration.
In the first example we assume a fixed external potential. 
In the second example we give an exactly solvable model including the
selfconsistent mean-field potential.

\subsection{Free Streaming}

The initial condition before folding is assumed as
\beq
  f_\delta(rp0)
  \propto
  e^{-\frac{\lambda}{2}p^2}\delta(r).
\eeq
The Vlasov-equation with $V=0$ yields then the time dependent solution
\begin{equation}\label{eq:micro}
  f_\delta(r,p,t)
  \propto
  e^{-\frac{\lambda}{2}p^2}\delta(r-{p \over m}t)
\end{equation}
or in vector notation ${\bf x}=(r,p)$
\begin{eqnarray}\label{eq:microvect}
  f_\delta(r,p,t)
  &=&
  \left(\lambda \over 2\pi\right)^{3/2}
  \left(\mu \over 2\pi \right)^{3/2}
  e^{-\frac{\lambda}{2}p^2-\frac{\mu}{2}(r-pt)^2}
\\  \nonumber
  &=&
  {  e^{-\frac{1}{2}{\bf x}^T\hat{{\bf \Lambda}}^{-1}{\bf x}}\over (2\pi )^{3} \sqrt{\mbox{Det}\{\hat{{\bf \Lambda}}\}}}
,\quad
  \mu
  \longrightarrow
  \infty
\end{eqnarray}
with
\begin{equation}\label{eq:deflam}
  \hat{{\bf \Lambda}}^{-1}
  =
  \left(\begin{array}{cc}
    \mu  & -\mu t \\
     -\mu  t  & {\lambda}+\mu  t^2
  \end{array}\right)
  \quad.
\end{equation}
The distribution (\ref{eq:microvect}) is to be folded with
\beq\label{eq:Husfold}
  {\cal G}(r,p)
&  =&
  {e^{-\frac{1}{2}{\bf x}^T\hat{{\bf \Sigma}}^{-1}{\bf x}}
\over (2\pi)^{3} \sqrt{\mbox{Det}\{\hat{{\bf \Sigma}}\}}}  
  \nonumber\\
  \hat{{\bf \Sigma}}^{-1}&=&\left(\begin{array}{cc}
    \sigma_r^{-2} & 0 \\
    0  & \sigma_p^{-2}
  \end{array}\right)
  \quad.
\eeq
Using the Gaussian folding theorem, we obtain
\begin{eqnarray}\label{eq:findis}
  f(r,p,t)
  &=&
  {  e^{-\frac{1}{2}{\bf x}^T\hat{\bf M}^{-1}{\bf x}}
\over (2\pi)^{3} \sqrt{\mbox{Det}\{\hat{\bf M}\}}}
  \quad,
\\[3pt]   \nonumber
  \hat {\bf M}
  &=&
 \hat {\bf \Lambda}+\hat {\bf \Sigma}
  =
  \left(\begin{array}{cc}
    \frac{1}{\mu }+\frac{t^2}{{\lambda}}+\sigma_r^{-2} & \frac{t}{{\lambda}} \\
    \frac{t}{{\lambda}}  & \frac{1}{\lambda}+\sigma_p^{-2}
  \end{array}\right)
\end{eqnarray}
and the entropy is
\beq\label{eq:entrop}
  S
&  =&
  -\int drdp\,f\log{(f)}
=
  \int drdp\,f(r,p,t)\nonumber\\
&\times&
\left(\frac{1}{2}{\bf x}\hat{\bf M}^{-1}{\bf x}
                    +3 \log{(2\pi)}
                    +\frac 1 2 \log{\left(\mbox{Det}\{\hat{\bf M}\}\right)}
              \right).
  \nonumber\\
&&
\eeq
Using known Gaussian integration rules
and performing the limit $\mu\longrightarrow\infty$, we obtain finally
\begin{equation}\label{eq:finentrop}
  S
  =
  3(\log{(2\pi)}+1)
  +\frac 3 2 \log{\left (\frac{1}{\sigma_r^2 \lambda}+\frac{1}{\sigma_r^2\sigma_p^2}+\frac{t}{\sigma_p^2 \lambda}\right )}
\end{equation}
which appreciates $S(t)=3 \log{\left ({2 \pi e  \over
      \sqrt{\lambda}\sigma_p} t\right )}$ for large times. Interestingly the long time limit is only modified by $\sigma_p$.
One can see that the entropy is monotonically increasing. The reason is the continuous solvation of the folded distribution.

\subsection{Selfconsistent bounded model}
\label{IIc}
After demonstrating the increase of entropy with time we like to consider two questions:
(i) Is this increase due to the
spreading of the Gaussian, which we assumed for space coarse
graining ? (ii) Can we interchange the
procedures {\it coarse graining} and {\it dynamical evolution}? This
means we like to check whether we will get identical results when we
first solve the exact Vlasov equation and then coarse grain the
solution or when we coarse grain the Vlasov equation first and solve
the modified one afterwards. This question reveals the sensible dependence on the initial distribution.

To answer both questions we consider another example of exactly
solvable model given in \cite{Mr97}, where we will add an external
harmonic oscillator potential. The potential $V(r,t)$  consists of the
separable multipole-multipole force $v_{1234}=v g_{12}g_{34}$ for the
mean-field and an external harmonic oscillator potential
\beq
V(\vec r,t)=v g(r) Q(t)+\frac 1 2 m \omega^2 r^2
\eeq
where
the selfconsistent requirement is
\beq
Q(t)=\int {d r d p\over (2\pi)^3} g(r) f_{\delta}(r,p,t)
\label{self}
\eeq
according to (\ref{vd}).
We may think of the external harmonic potential as a representation of a realistic confining potential expanded around the origin $U_{\rm ext}(r)\approx U_{\rm ext}(0) (1-(r/R)^2/2)$ with the radius $R$. We have therefore
\beq
\omega^2\approx -U_{\rm ext}(0)/m R^2.\label{om}
\eeq
The one-particle distribution function obeys the quasi-classical Vlasov equation (\ref{vlas}). We can solve the Vlasov equation exactly by solving the differential equation for the equipotential lines, which are the Hamilton equations. We choose a form factor of
\beq
g(r)=a_x r_x +a_y r_y+ a_z r_z.
\label{pot}
\eeq
This model is special by two reasons. Firstly, for this model the linearization (\ref{coll}) is exact, because higher than second order spatial derivatives vanish. Secondly, within this model the quantum-Vlasov equation agrees with the semiclassical Vlasov equation investigated here.

The Hamilton equations of trajectories which correspond to this model
\beq
\partial_t p&=&-v a Q(t) -m \omega^2 r\nonumber\\
\partial_t r&=&{p \over m}
\label{hamilt}
\eeq
are solved as
\beq
\left (\matrix{r\cr p}\right )&=& \left(\matrix{\sin{\omega t} & \cos{\omega t}\cr
m \omega \cos{\omega t}&-m\omega \sin{\omega t}}\right) \left(\matrix{c_1\cr c_2}\right)
\nonumber\\
&-&v a\int\limits_0^{t} dt' Q(t') \left(\matrix{{1 \over m\omega} \sin{\omega (t-t')}\cr\cos{\omega(t-t')}}\right).\label{coord}
\eeq
Now we know that the constants of motion $c_1,c_2$ are constant at any time of evolution. Therefore we can relate them to the initial momenta and positions which gives
\beq
\left(\matrix{c_1\cr c_2}\right)=\left(\matrix{\frac{p_0}{m\omega}\cr r_0}\right).
\eeq
From equation (\ref{coord}) we express now $r_0$ and $p_0$ as functions of $(r,p,t)$ which results into
\beq
\left (\matrix{r_0\cr p_0}\right )&=& \left(\matrix{\cos{\omega t} & -{1\over m \omega}\sin{\omega t}\cr m \omega \sin{\omega t}&\cos{\omega t}}\right)
\nonumber\\
&\times&\left( \left(\matrix{r\cr p}\right)+\mu a\int\limits_0^{t} dt' Q(t') \left(\matrix{{1 \over m\omega} \sin{\omega (t-t')}\cr\cos{\omega(t-t')}}\right)\right )\nonumber\\
&\equiv&\hat {\bf A} \left (\matrix{r\cr p}\right )+{\bf C}.
\label{coord1}
\eeq
Given an initial distribution $f_0(r_0,p_0)$, we can express the
solution at any time as
\beq
f_{\delta}(r,p,t)&=&f_0(r_0,p_0)
=f_0(\hat {\bf A} {\bf x}+{\bf C})
\label{sol}
\eeq
with ${\bf x}=(r,p)$ and the matrix $\hat {\bf A}$ and the vector ${\bf C}$ can be read off from (\ref{coord1}).
The selfconsistency requirement (\ref{self}) leads to a determination of Q(t)
\beq
Q(t)&=&a <r_0> {\rm cosh}\sqrt{\Omega^2-\omega^2} \,t
\nonumber\\
&+&{a \over m\omega\sqrt{\Omega^2-\omega^2}}<{p \over m}>{\rm sinh}\sqrt{\Omega^2-\omega^2} \, t
\label{se}
\eeq
where
\beq
\Omega^2&=&-{v a^2\over m} N\nonumber\\
N&=&\int {drdp \over (2\pi \hbar)^3} f_0(r,p)\nonumber\\
< r_0>&=&\int {drdp \over (2\pi \hbar)^3} r f_0(r,p)\nonumber\\
<{ p_0 \over m}>&=&\int {drdp \over (2\pi \hbar)^3} { p \over m}
f_0(r,p)\nonumber\\
\eeq
are expressed by moments of the initial distribution.

We have given an exact solution of the selfconsistent 3-D Vlasov equation. It is interesting that for $\Omega>\omega$ we have a positive Lyapunov exponent $\sqrt{\Omega^2-\omega^2}$ while in the opposite case the solution oscillates. 
We like to point out that (\ref{sol}) represents a general solution of any Vlasov equation if we understand $\hat {\bf A} {\bf x} +{\bf C}$ as a nonlinear transformation solving the corresponding Hamilton equations.


In analogy to the foregoing chapter we can now coarse grain this exact
solution. We are able to check by this way whether the so far observed
entropy increase is due to the spreading of the test particles with
Gaussian width. If so, we would observe an entropy increase even if we
start with an equilibrium solution. We will demonstrate now that this
is not the case. Instead, we have a nonlinear mode coupling. The
actual calculation is performed in the appendix \ref{melfc}. We choose
as an initial distribution a Gaussian one.
The entropy finally reads from (\ref{entra})
\beq
  &&S
  =
  3\log{(2\pi e) }
  +\frac 3 2 \log
\left \{
\frac 1 2 (\sigma_p^2+m^2 \omega^2 \sigma_r^2)({1\over \lambda m^2\omega^2}+{1\over \mu})\right .\nonumber\\
&&\left . +
({1 \over \lambda \mu}\!+\!\sigma_p \sigma_r)\!+\!
\frac 1 2 (m^2 \omega^2 \sigma_r^2\!-\!\sigma_p^2)({1\over \lambda m^2\omega^2}\!-\!{1\over \mu})\cos(2 \omega t)
\right \}.\nonumber\\&&
\label{entrb}
\eeq
The entropy is just oscillating around a stationary value. The Gaussian form of initial distributions leads to {\it no} stationary solution. This shows that the entropy increase observed so far is not due to the spreading of Gaussians but due to the form of initial distribution. Other initial distributions would lead to an increase of entropy which can be checked e.g. with ground state Fermi functions.

Now we turn to the second question concerning the interchange between
{\it coarse graining} and {\it solution of kinetic equation}. While we
have solved the exact Vlasov equation so far and coarse grained the
solution afterwards we now revert the procedure and solve the coarse
grained Vlasov equation (\ref{vlas1}) with (\ref{coll}). Please remind
that the linearization with respect to the width is exact for this
model here.
The entropy in this case is calculated from (\ref{solb}) with the result
\beq
  &&S
  =
  3\log{2 \pi e }
  +\frac 3 2 \log
\left \{ {1\over \lambda \mu}+\frac{1}{2}
(\sigma_p^2-m^2 \omega^2 \sigma_r^2) \right .\nonumber\\
&&\times \left . ({1\over \lambda m^2\omega^2} -{1\over \mu}-{\sigma_p^2\over m^2 \omega^2}+\sigma_r)
(1-\cos(2 \omega t))
\right \}.
\eeq
Comparing with (\ref{entrb}) we see a different expression, however the oscillating behavior remains. Within linear orders of $\sigma$ both expressions differ by a constant of $2{\sigma_p^2 \over \mu}+2 {\sigma_r^2  \over \lambda}$. The difference between these two expressions, which corresponds to the interchange of coarse graining and dynamics is explained by the use of the same initial distribution in both cases. We obtain a different dynamical behavior. If we would additionally coarse grain the initial distribution (\ref{ftfo}) we would have to replace $\hat {\bf \Lambda}$ by $\hat {\bf \Lambda} +\hat {\bf \Sigma}^{-1}$.  Instead of (\ref{M}) we would obtain
$\hat M=(\hat {\bf A}^T (\hat {\bf \Lambda} +\hat {\bf \Sigma}^{-1})\hat {\bf A})^{-1} -\hat {\bf B}
$
and the resulting entropy agrees exactly with (\ref{entrb}). As already
pointed out this would be not the case for nonlinear models other than
(\ref{pot}) since then the selfconsistent condition (\ref{self}) will be
affected by coarse graining itself.

\subsection{Consequences}

This observation has some practical consequences. Since in all
numerical calculations one solves the coarse grained Vlasov equation
instead of the exact one, which corresponds to first coarse graining
and then solving, one should not expect the correct dynamical
behavior starting from a fixed initial distribution. Instead the
correct procedure is to coarse grain first the initial distribution
with the Gaussians of the used test-particles and then solve numerical
the time evolution of this distribution. The refolding at any time
step yields then the exact solution of Vlasov equation for such
trivial models as described here where the selfconsistency condition
is not altered by the coarse graining. 

It is important to stress again that the above equivalency between the two ways
of solving and coarse graining are only equivalent in the linear model
(\ref{pot}) which we have solved exactly. 
For more realistic potentials the selfconsistency condition like
(\ref{self}) becomes altered itself by the coarse graining which leads
to an essential nontrivial change in the dynamics. Therefore both
method will lead to different results. The nonlinear dynamics is
non-trivially changed if one coarse-grains first and solves then or if
one solves first and coarse-grains afterwards. However, in order to diminish the numerical error due to coarse graining the refolding should be performed at least.

\subsubsection{Time scale of entropy production}

From the chapter above we see that the entropy production due to coarse graining in an harmonic external potential is oscillating like $1-\cos{(2 \omega t)}$. We can use this fact to define the typical time scale of entropy production
\beq
\tau_c={\pi \over 2 \omega}\approx {\pi R \over 2}\sqrt{{m \over U_{\rm ext}(0)}}
\eeq
where we used (\ref{om}). For a typical nuclear situation we obtain therefore $\tau_c\approx 4.8 A^{1/3}$fm/c such that for $^{12}O$ we get $\tau_c\approx 12$fm/c and for $^{208}Pb$ we have $\tau_c\approx 28$fm/c.
It is remarkable that the time scale on which the entropy is changing is independent of the used internal interaction. This clearly underlines the Landau damping type of dissipation. Coarse graining produces no genuine dissipation.

\section{Stationary solution of coarse grained Vlasov equation}

We will demonstrate in this chapter that the solution of the general coarse grained Vlasov
equation is approaching a modified Boltzmann limit for long times
provided a stationary solution is approached at all. The example in
chapter \ref{IIc} has shown that this is not the case every
time. Exclusively for appropriate initial conditions, which will be
characterized in chapter \ref{stab}, we will obtain stationary
solutions due
to the additional terms in the Vlasov equation
(\ref{vlas1}).

\subsection{Solution of coarse grained Vlasov equation}

We want to consider the solution of the kinetic equation (\ref{vlas1}) with
the assumption of arbitrary time dependent mean-fields. Therefore we solve the
following partial differential equation neglecting the selfconsistency in $V$
at first and remember it later. We consider
\begin{eqnarray}
  {\pa t} f(prt) &+&\Phi(r,t) {\pa r}{\pa p} f(prt) +{p \over m} {\pa r} f(prt)
\nonumber\\
 &-&{\pa r} V(r,t)
{\pa p}  f(prt)=0
\label{vlas2}
\end{eqnarray}
with $\Phi(r,t)=({\sigma_p^2 \over
      m}-\sigma_r^2{\partial^2 \over \partial r^2} V(r,t))$ and the boundary conditions
$f(r,p,0)=f_0(rp), \int dp dr f(rpt)/(2 \pi \hbar)^3=N$.

This equation is of parabolic type and as an initial value problem well defined with an unique
solution.
The unique stationary solution, if such a stationary solution is approached, is
an implicit representation of the stationary solution
by the remembrance of the dependence of $V$ on the distribution function itself, i.e. the selfconsistency. Because
the problem is uniquely defined it is enough to find a special representation of the solution.

\subsubsection{Representation of stationary solution}

The stationary solution of (\ref{vlas2}) can be found by separation of
variables. Assuming $f_{\rm stat}(pr)=f_P(p)f_R(r)$ we have

\begin{equation}
  f_{\rm stat}^n(pr)={\rm const} \times {\rm exp} \left [-c_n({p^2 \over 2m}+\int\limits^r dr' {\partial_r'
    V(r') \over 1 - c_n \Phi(r')})\right ],\label{solut}
\end{equation}
with the separation constant $c_n$ which holds for vectors $p,r$.
The general stationary solution is given by superposition of these $c_n$ dependent expression (\ref{solut}). Linearizing via the coarse graining $\sigma$ leads us to
\beq
f_{\rm stat}(p,r)&=&\sum\limits_n a_n {\rm e}^{-c_n ({p^2 \over 2 m}+V(r))} \nonumber\\
&\times&(1- c_n^2 ({\sigma_p^2 \over m} V(r) -{\sigma_r^2 \over 2}(V'(r))^2 ) )\label{soluta}
\eeq
from which we deduce the equilibrium density distribution to have the form
\beq
n(r)&=&\sum\limits_n a_n ({m \over 2 \pi \hbar^2 c_n})^{3/2} {\rm e}^{-c_n V(r)} \nonumber\\
&\times&(1- c_n^2 ({\sigma_p^2 \over m} V(r) -{\sigma_r^2 \over 2}(V'(r))^2 ) ).
\eeq
Please remark that the summation can also be replaced by an
integration which would translate into continuous functions $c_n$
and $a_n$. For the reason of legibility we discuss further on only the
discrete case.

With this solution we have derived our goal to present a stationary solution
of the modified Vlasov equation. Because the initial problem was uniquely
defined this solution is the unique stationary one. It has to be remarked that
due to the selfconsistency ( dependence of $V$ on the distribution
function itself ) equation (\ref{soluta}) is an implicit representation of the solution.

\subsection{Determination of $c_n$}
\label{cn}
The open expansion coefficients $c_n$ are completely determined by the
initial distributions. This can be seen as follows. To solve the
Vlasov equation we can rewrite the solution into Hamilton equations
with a time dependent potential. This time dependence comes from the
selfconsistent potential and/or from an external potential. The
selfconsistency is then represented by a nonlinear determining
equation similar to (\ref{self}). The Hamilton equation for the
trajectories can be solved in principle with integration constants
$c_i$ as we demonstrated in the model. Because these are constants at
any time we obtain a
transformation between initial coordinates and the coordinates at a
later time, including the time dependence as $(r_0,p_0)=A[r,p,t]$. Since
we like to solve the Vlasov equation as initial value problem with
a given initial distribution $f_0(r_0,p_0)$, the time dependent solution
is given formally by $f(r,p,t)=f_0(A[r,p,t])$. {\it If} this solution approaches a stationary state,
which is dependent on the initial distribution as well as on the interaction potential, we obtain with the result of the foregoing chapter
\beq
f_0(A[r,p,\infty])&=&\sum\limits_n a_n {\rm e}^{-c_n ({p^2 \over 2 m}+V(r))} \nonumber\\
&\times&(1- c_n^2 ({\sigma_p^2 \over m} V(r) -{\sigma_r^2 \over 2}(V'(r))^2 ) ).\label{56}
\eeq
By integrating over $r$ and inverse Laplace transform it is possible
to extract the $c_n$ uniquely. In the next chapter we will find that
only such initial distributions will lead to stationary solutions
which obeys $c_n<{m\over 2 \sigma_p^2}$. Since the $c_n$ are
determined by the initial distribution and the used potential, we can
decide which initial distributions will lead to stationary solutions
with a given interaction potential and coarse graining $\sigma_p$.

\subsection{Global stability of the stationary solution}
\label{stab}
Since the general possible solution of the stationary coarse grained
Vlasov equation covers a range of unphysical solutions which will
never be reached, we have to ask which solutions are the stable
ones. Therefore we employ a linear response analysis of the Vlasov
equation (\ref{vlas2}) to an external potential $U_{ext}$. We assume a homogeneous equilibrium characterized by the distribution $f_0(p)$. Then the equation (\ref{vlas2}) can be linearized according to $f(prt)=f_0(p)+\delta f(prt)$ as
\beq
(-i \omega + i {p k \over m}) \delta f(pk\omega) &=& i V'(n_0) k \partial_p f_0  \delta n(k,\omega) \nonumber\\
&+& i k \partial_p f_0 U_{ex}(k\omega)\nonumber\\
&-&i {\sigma_p^2 \over m} k \partial_p \delta f(pk\omega).
\eeq
Here we have Fourier transformed $r, t$ coordinates. It can be easily solved for $\delta f$. Integrating this solution over $p$ an algebraic solution for $\delta n$ is obtained

\beq
\delta n = U_{ex} {\Pi(k\omega) \over 1- V'(n_0) \Pi(k\omega)}
\label{dn}
\eeq
with the polarization function
\beq
\Pi(k\omega)&=&\int {dp \over (2 \pi \hbar)^3}(1- {\sigma_p^2 \over m} {k \partial p \over \omega-{p k\over m}}) {k \partial_p f_0 \over \omega-{p k \over m}}\nonumber\\
&=&\int {dp \over (2 \pi \hbar)^3}(1+ {k^2 \sigma_p^2 \over m^2} {1 \over (\omega-{p k\over m})^2}) {k \partial_p f_0 \over \omega-{p k \over m}}.
\label{Pi}
\eeq
We see that the usual RPA response function is modified by a structure function
\beq
M^2(kp\omega)=1+{k^2 \sigma_p^2 \over m^2} {1 \over (\omega-{p k\over m})^2}.
\label{appr}
\eeq
This describes the fact that the elementary particle considered here
(testparticles) have a finite width or an internal structure. For
large distances $k\rightarrow 0$ we see that $M$ approaches 1, which
means that this structure is only resolvable at smaller distances.

A known approach to find the structure functions ${\tilde M}$ inside the RPA polarization function is to include vertex corrections. The resulting response functions can be written generally
\beq
\Pi(k \omega)=\int {dp \over (2 \pi \hbar)^3}
{\tilde M}^2(kp\omega) {k \partial_p f_0 \over \omega-{p k \over m}}.
\eeq
By approximating this structure function $\tilde M$ by (\ref{appr}) we might simulate higher order correlations by finite momentum widths of testparticles.
Also the influence of finite size can be compared with this expression
\cite{MVFLS98,M99}.

The equilibrium solution is stable as long as the increment of the collective mode does not change the sign. Otherwise the collective mode would exponentially increase with time and the solution would be unstable with respect to small perturbations. The collective mode is given in linear response by the complex solution of $1-V'(n_0) \Pi(k\omega)=0$ of eq. (\ref{dn}). For small increments the sign of the complex part is completely determined by the sign of ${\rm Im} \Pi$ of (\ref{Pi}). Let us check this stability condition for the stationary solution.
From (\ref{solut}) we have

\beq
f_{\rm stat}(p)&=&\int dr f_{\rm stat}(p,r)=\sum\limits_n a_n {\rm e}^{-c_n ({p^2 \over 2 m})} d_n
\eeq
where $d_n$ is the spatial integral over the potential dependent part.
From (\ref{Pi}) we obtain as criterion of stability
\beq
&&{\rm Im} \Pi=-(1+{k^2\sigma_p^2\over m^2}{\partial^2\over \partial \omega^2}){m^2 \omega \over 2 \pi k}\sum\limits_n a_n d_n {\rm e}^{-{c_n\over 2 m} \left ({m \omega\over k}\right )^2}\ge 0.\nonumber\\&&\label{ssum}
\eeq
To provide stable solutions we demand that any term in the sum should
be positive. We observe now two sources of possible instability: (i)
The expansion coefficient $a_n$ without coarse graining can become
negative for enough pathologic initial functions. For such initial distributions we would not get stable solutions by Vlasov dynamics itself. (ii) Provided the original distribution is stable $a_n>0$ we find an additional criterion for stability if we use coarse graining. Demanding the coefficients to be positive
leads us from (\ref{Pi}) to the equation
\beq
1+\sigma_p^2 (-{3 c_n\over m}  +c_n^2 ({\omega \over k})^2)\ge 0.
\eeq
In the case of small wave lengths ${m\omega \over k}\rightarrow \infty$ we obtain the most restrictive condition 
\footnote{The less restrictive condition for stability reads
\beq
c_n\not\in {3 k^2\over 2 m \omega^2}\left (1- \sqrt{1-{4 m^2\omega^2\over 9 k^2 \sigma_p^2}},1+ \sqrt{1-{4 m^2 \omega^2\over 9 k^2 \sigma_p^2}}\right ).
\label{co}
\eeq
In the general case with arbitrary wave length we see that (\ref{co}) is fulfilled if
\beq
\sigma_p<\frac 2 3 {m |\omega| \over k}.
\label{c2}
\eeq
We conclude that all solutions (\ref{solut}) are stable if the coarse graining is smaller than the typical scales in the system (\ref{c2}). If (\ref{c2}) is not fulfilled, than only such solutions (\ref{solut}) are stable solutions which expansion coefficients $c_n$ are smaller than the inverse coarse graining via (\ref{c1}). Therefore (\ref{c1}) is the most restrictive condition.}
\beq
c_n<{m\over 2 \sigma_p^2}.
\label{c1}
\eeq
If this condition is not fulfilled we find $m \omega/k$ combinations such that the sum in (\ref{ssum}) can change the sign and the solution is unstable. In other words (\ref{c1}) is a necessary condition for stability of the stationary solution.
Since the $c_n$ are determined by the initial distribution according
to (\ref{56}) and the potential as discussed in chapter \ref{cn}, we
see that only special initial distributions can lead to stationary long time distributions due to coarse graining with a given potential.

Interestingly, for a Maxwellian initial distribution $c_n=1/k_B T$ and $a_n={\rm const} \delta_{n,0} $, the coarse graining has to obey
\beq
\sigma_p^2<{m \over 2} k_B T
\eeq
to provide stable solutions. This gives the intuitive clear
interpretation that the coarse graining energy $\sigma_p^2/2 m$ should
be smaller than a quarter of the kinetic energy in order to render
numerical investigations stable.

\section{Thermodynamic consequences}

We have reached our goal to show that phase space coarse graining leads to
dissipative features of reversible kinetic equations and the special coarse
graining with Gaussian width forces the system to a sum of modified
Boltzmann - like
distributions. We have shown algebraically the numerical observed fact that in ordinary Boltzmann codes two different limiting
values of the distribution functions are appreciated \cite{RS95b}. From one-
particle dynamics a Boltzmann- like distribution and from quantum
Boltzmann collision
integrals a Fermi distribution evolves. In contrast to earlier work we
find that even for an infinite number of test-particles the Boltzmann limit is
appreciated due to the finite width of test-particles.

From the found stationary solution of the coarse grained Vlasov equation we
can now derive thermodynamic consequences.
We assume for simplicity a Maxwellian initial distribution which may
evolve in time via the coarse grained Vlasov equation. Then the
coefficients $c_n$ of the stationary solution (\ref{soluta}) are only single parameters
determined by $c_n=\beta=1/k_B T$ and
\beq
&&a_n=\delta_{n,0} {{N \lambda_T^3 \over g} \over \int d r {\rm e}^{-\beta V(r)}
\left \{1-\beta^2 [V(r)
  {\sigma_p^2 \over m}- {\sigma_r^2 \over 2} (\partial_r V(r))^2]\right \}} 
\nonumber\\&&
\label{64}
\eeq
by the requirement of particle conservation that the integral
over momentum and space of (\ref{soluta}) should equal to the particle number $N$. Here $g$ describes the spin-isospin degeneracy and $\lambda_T^2=2 \pi \hbar^2/m k_B T$ and we use the linearization in $\sigma$ since the
original equation (\ref{vlas2}) is valid only up to orders of $o(\sigma^4)$.

From the expression (\ref{soluta}) we see that the coarse graining leads to a
modification factor of the distribution function $f_{\rm stat}(pr)$ in comparison with the Maxwell-Boltzmann distribution function $f_M(pr)={\rm const \,exp} (-\beta p^2/2m -\beta V(r))$
\beq
&&{f_{\rm stat}(pr) \over f_M(pr)} =
{1-\beta^2 \left (V(r){\sigma_p^2 \over m}-
{\sigma_r^2 \over 2}(\partial_r V(r))^2\right )
\over 1-\beta^2 <
\left (V(r){\sigma_p^2 \over m}-
{\sigma_r^2 \over 2}(\partial_r V(r))^2\right )>
}
\nonumber\\
&=&1-\beta^2 \left ({\sigma_p^2 \over m}\left [V(r)-<V(r)>\right ] \right .
\nonumber\\
&&\left . \qquad \qquad -
{\sigma_r^2 \over 2}\left [(\partial_r V(r))^2-<(\partial_r V(r))^2>\right ]\right )
\label{thermo}.\nonumber\\&&
\eeq
where the spatial average is abbreviated as\\
$<a>=\int dr {\rm e}^{-\beta V(r)} a/\int dr {\rm e}^{-\beta V(r)}$.

As a consequence, no contribution of coarse graining occurs for mean
values of momentum dependent observables within the lowest order of coarse gaining width. However, for space dependent observables one gets a modified thermodynamics by $\sigma_p$ coupling the mean spatial fluctuation of the potential and by $\sigma_r$ coupling the fluctuation of the potential derivative.
Consequently, only the {\it spatial} thermodynamical quantities will be influenced by the underlying
coarse graining. We can consider this modification factor as the
expression of spatial correlations induced by momentum and spatial coarse graining.

\begin{figure}
\psfig{file=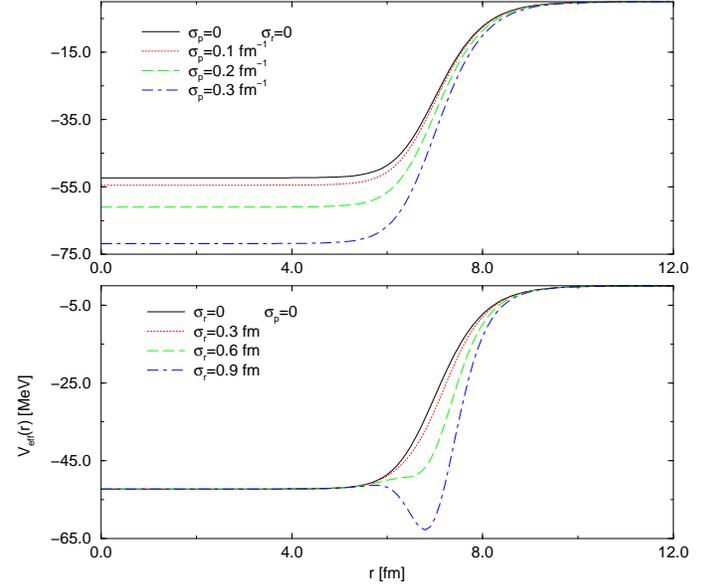,width=9cm,angle=-90}
\caption{The influence of different coarse graining parameters on
the selfconsistent potential. Above the space coarse graining
is zero and below the momentum coarse graining is set to zero.
While the momentum coarse graining leads to an increase of
the selfconsistent potential the spatial coarse graining
enhances the gradient and produces a skin.}
\label{tf}
\end{figure}

\subsection{Selfconsistency}

Now we return to the question of selfconsistency. We have so far
assumed silently that (\ref{vd}) or (\ref{vr}) are completed by a
selfconsistent potential $V(r)$. Without coarse graining this would be
accomplished by solving the Thomas--Fermi--like equation (\ref{vd})
with $f\propto \exp{-\beta (p^2/2m+V(r))}$
\beq
V(r)={N \over g}{\int dr'  V_0(r-r')
{\rm e}^{-\beta V(r')}\over \int dr'
{\rm e}^{-\beta V(r')}}
\label{tf1}
\eeq
and analogously for the degenerate case.
For the coarse grained case we see now from (\ref{soluta}) and
(\ref{64}) that the potential in
the distribution function $f$ has to be replaced by 
\beq
V_{\rm eff}(r)=V(r)+\beta \left [ {\sigma_p^2 \over m}V(r)-{\sigma_r^2
\over 2} (\partial_r V(r))^2 \right ].
\label{pots}
\eeq
Therefore one can solve the Thomas Fermi like equation (\ref{tf1}) as
before but replace afterwards $V\rightarrow V_{\rm eff}$.
In the following let us assume we have solved the Thomas--Fermi equation
(\ref{tf1}) and we will discuss now the influence of coarse graining.

We see that the coarse graining in momentum space increases the
potential (\ref{pots}) globally. In contrast, the coarse graining in
space 
causes a
widening of the potential. This is demonstrated in figure
\ref{tf}, where we assume a Wood--Saxon potential for $V(r)$ in $^{208}Pb$ 
and plotted
the change of the potential for
different parameters of coarse graining
at a temperature of $10$ MeV. While the momentum coarse
graining leads to an overall increase, the spatial graining
sharpens the gradient in the potential and causes the appearance
of a skin.

One can consider this as a modification of the Thomas- Fermi equation by
the finite width of phase space graining. The coarse graining obviously
produces higher binding properties and larger rms- radii. The appearance of
a skin is important to note as a relict of the testparticle width.

\subsection{Consequences on collective modes}

As a practical application we show now that the coarse graining which is numerically unavoidable leads to
false predictions concerning the energy and the width of giant resonances. For illustrative purpose we restrict to the low temperature $T=0$ case. The collective mode is given by the complex zeros $\omega=E-i\Gamma$ of the denominator of (\ref{dn}). This solution provides us with the centroid energy $E$ and damping $\Gamma$ of the collective mode. We find for $T=0$ the dispersion relation
\beq
&&1-V \Pi(q,\omega)\equiv 1+F_0 (1+{\sigma_p^2\over 2 p_f^2 }{\partial^2\over \partial u^2})\nonumber\\
&&\times
\left [-1+{u \over 2} {\rm ln}\left ({1+u \over 1-u}\right)-i{\pi \over 2}u \Theta(1-u) \right ]=0\label{lint0}
\eeq
where we have introduced $u=m\omega/p_f/k$, the Landau parameter $F_0={2 V'(n_0) m p_f \over \pi^2 \hbar^3}$ and the Fermi momentum $p_f$.
\begin{figure}
\psfig{file=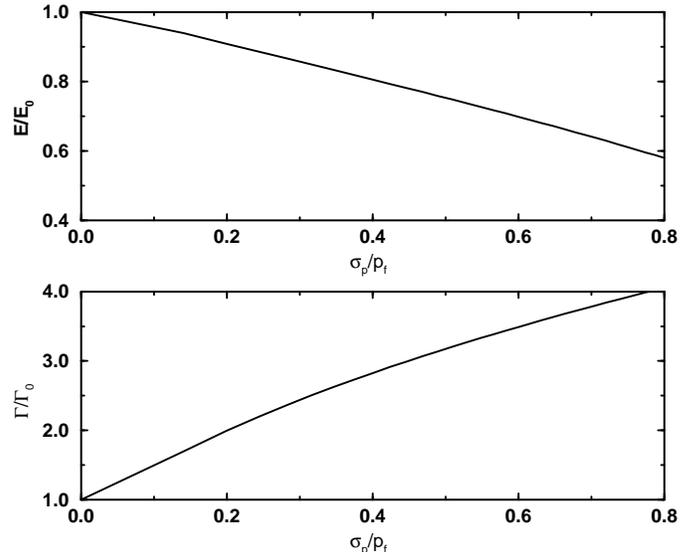,width=9cm}
\caption{The centroid energy and width of collective motion versus the pseudoparticle width in the low-temperature linear regime (\protect\ref{lint0}). The values are scaled to the free one $\sigma_p=0$.}
\label{width}
\end{figure}

In Figure \ref{width} we give the ratio of the centroid energy and damping to the corresponding un-coarse grained ones for typical nuclear situation. We see that with increasing width of coarse graining the centroid energy decreases and the width increases. This is understandable because due to coarse graining we lower the particle- hole threshold resulting into lower centroid energy and an artificial damping at the same time. Since coarse graining is unavoidable in numerical implementations the extraction of damping width of giant resonances should be critically revised with respect to the used pseudoparticle width.

\begin{figure}
\psfig{file=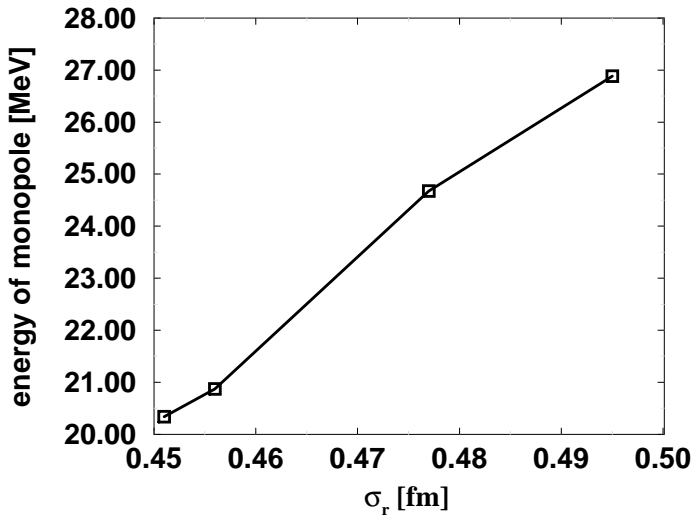,width=9cm}
\caption{The centroid energy of monopole resonance in $^{40}Ca$ versus spatial
testparticle width within a Vlasov simulation. The centroid energy is increasing with increasing width.}
\label{ca}
\end{figure}

In contrast to the coarse graining in momentum space, we find a different behavior in spatial domain.
In figure \ref{ca} we plot a realistic numerical solution of the full Vlasov equation
describing monopole resonances in $^{40}Ca$. Here the spatial width of the test particles
has been varied. We see that the centroid energy is increasing with increasing testparticle width.
This behavior can be understood from figure \ref{tf}. The spatial coarse graining enlarges the gradient of the potential and therefore increases the restoring force. This translates into a higher incompressibility and a higher collective energy.

A more complete discussion of the solution of Vlasov equation and
also application to giant resonances far from the stability line can be
found in \cite{MFW00}.

\section{Summary}

The dissipative features of coarse grained Vlasov equations are investigated.
This procedure occurs due to numerical simulation techniques. We
have calculated explicitly the entropy production, which is due to nonlinear mode coupling but not due to
dissipation. We find that a sum of modified Boltzmann distributions is approached by the coarse graining. Examples are shown where no stationary condition is
approached at all since the solution oscillates and examples are given where
a stationary solution can be reached. The different behavior is completely
determined by the initial distribution and the used potential. The stability
analysis leads to a criterion for the initial distribution dependent on
the width of test particles, which can only lead to stationary solutions.

We have demonstrated by a special model that the two steps, coarse graining and dynamical evolution of
the distribution function, are only interchangeable if the initial distribution
is also coarse grained. It is argued that this property does not hold in general due to the feedback of the selfconsistent potential. Because the coarse graining is unavoidable
in numerically implementations, the Vlasov codes should be critically
revised with respect to the question if they start really from a coarse
grained initial distribution.

Thermodynamical consequences are discussed. A correlated part to any
thermodynamical observable is calculated explicitly. It is given in terms
of the space and momentum width. We find that the selfconsistently
determined nuclear potential is overestimated by testparticle simulations with
finite width of the test particles. This should have
implementations on Thomas- Fermi calculations where an overbinding is found.

The linear response function is calculated for homogeneous systems and the
spectra of density fluctuation is presented. It is found that the RPA
polarization function becomes modified due to the finite momentum width of
testparticles. These modifications can be understood as an internal structure
the particles bear. This structure function is formally compared with
vertex corrections to the RPA. It is pointed out that the higher order
vertex corrections beyond RPA can be casted into similar structure functions.
Therefore we suggest a method of simulating higher order correlations in
one-body treatments by choosing an appropriate momentum width of testparticles.

As a practical consequence the collective mode is analyzed. We find that the
coarse graining enhances the damping width and lowers the centroid energy of
collective modes, e.g. giant resonances. The dependence on the coarse graining
width is given quantitatively and corresponding simulations should be revised.

\section{Acknowledgments}

The authors are especially indebted to P.G. Reinhard for many
discussions and critical comments. The model IIIA has been contributed
by him. J. Dorignac is thanked for critical reading and helpful
comments. The friendly and hospitable atmosphere of LPC in Caen is
gratefully acknowledged.

\appendix

\section{Refined testparticle picture}\label{app}

We want to consider now the question how the dynamics is influenced by coarse graining. While the original Vlasov equation (\ref{vlas}) is represented by the Hamilton dynamics of testparticles via (\ref{sum}), we like to know how the equation of motions are changed by coarse graining. Therefore we obtain with (\ref{rel}) the coarse grained kinetic energy $K$ and the potential energy $W$ as
\beq
K(t)&=&\int {dp dr \over (2\pi \hbar)^3} {p^2 \over 2 m} f_{\delta}(p,r,t)\nonumber\\
&\equiv& \int {dp dr\over (2\pi \hbar)^3}  {p^2 +\sigma_p^2 \over 2 m} f(p,r,t)\nonumber\\
W(t)&=&\frac 1 2 \int {dp dr \over (2\pi \hbar)^3}  V_{\delta}(r,t) f_{\delta}(p,r,t)
\nonumber\\
&\equiv& \frac 1 2 \int {dp dr\over (2\pi \hbar)^3}  V(r,t)  \rm{exp}(\sigma_r^2 {\leftvector_r} {\rightvector_r}) f(p,r,t).\label{prin}
\eeq
We see that the coarse graining in momentum space leads to an additional apparent temperature $\sigma_p^2=m T$, while the coarse graining in space modifies the potential and therefore the dynamics.

With the help of (\ref{sum}) and (\ref{f}) we find from the variation of the
action $\int dt (K(t)-W(t))$ just the Hamilton equations
\beq
{\pa t} R_i(t) &=& {P_i(t) \over m}\nonumber\\
{\pa t} P_i(t) &=& -{\pa R_i} V_{\delta}(R_i,t)
\eeq
irrespective which form we use of (\ref{prin}). This is clear because the underlying dynamics is entirely covered by the Hamiltonian dynamics. Therefore even the center of mass movement of the Gau\ss{} packets $R_i,P_i$ follows the Hamilton dynamics.

We may now turn the question around and define a new quasi(test)- particle picture. We seek the equation of motion for new testparticles $(\tilde P_i,\tilde R_i)$ which should now represent the {\it coarse grained } distribution (\ref{f})
\beq
  f(prt)&=&\{f_{\delta}\}_g\nonumber\\
&=&\sum\limits_{i=1}^{\infty} g_r(r-R_i(t))g_p(p-P_i(t))\nonumber\\
&\equiv&\sum\limits_{i=1}^{\infty} \delta (r-\tilde R_i(t))\delta (p-\tilde P_i(t)).\label{f1}
\eeq
The last step is just the definition of the new test particles which
obey the equation of motion derived from (\ref{prin})
\beq
{\pa t} \tilde R_i(t) &=& {\tilde P_i(t) \over m}\nonumber\\
{\pa t} \tilde P_i(t) &=& -{\pa \tilde R_i} {\rm e}^{-\sigma_r^2 \partial_{\tilde R_i}^2} V (\tilde R_i,t)
\eeq
with $V$ the coarse grained mean field (\ref{vr}). We observe that the coarse grained distribution $f$ can be represented by a set of quasi(test)- particles which obey Hamilton equations with a modified effective potential
\beq
V_{\rm eff}(r,t)&=&{\rm e}^{-\sigma_r^2 \partial_{r}^2}V(r,t)\nonumber\\
&\approx&V(r,t)-\sigma_r^2 \partial_r^2 V(r,t).
\eeq
We see that the relation between the coarse grained mean field potential (\ref{vr}) and the effective one can be written as a double folding
\beq
V(r,t)&=&{\rm e}^{\sigma_r^2 \partial_{r}^2} V_{\rm eff}(r,t)\equiv\{\{V_{\rm eff}(r,t)\}_{g_r}\}_{g_r}
\eeq
or
\beq
V_{\delta}(r,t)&=&\{V_{\rm eff}(r,t)\}_{g_r}\label{pote}
\eeq
where we have used the relation $\{f(r)\}_{g_r}=\exp{({\sigma^2 \over 2} \partial_r^2)} f(r)$, see \cite{OWW84}. Please observe that the here presented relation is just the inverse relation given in \cite{RS95a,RS95b,LSR95}. The difference comes from the inverse picture used here. The authors of \cite{RS95a,RS95b,LSR95} investigated how the dynamical equation change {\it if in} the action the distribution functions are replaced by their coarse grained ones. We present here the opposite view that the action is unchanged but is rewritten according to coarse graining, leading to modified {\rm explicit} kinetic and potential energy (\ref{prin}). This still leads to unmodified equation of motions for the center of mass coordinates of the Gau\ss{} packets as outlined above. When we now represent the coarse grained distribution by a new set of sharp testparticles, the dynamics becomes modified and an effective potential appears which is the {\it inverse} folded mean field 
potential (\ref{pote}). 

We conclude that we can map the coarse grained Vlasov equation to the original Vlasov equation if testparticles are introduced which obey a modified dynamical equation with the new inverse folded mean field potential. Within the text we have followed the other route to describe the influence on the dynamics in the old picture to make the effects more transparent.

\section{Coarse graining of selfconsistent model}\label{melfc}

Here we give the explicit calculations of chapter \ref{IIc}.

\subsubsection{First solving than coarse graining}

We like to choose as an initial condition a Gaussian distribution, which represents an equilibrium distribution
\begin{eqnarray}
  f_0({\bf x})
  &=&
  \left(2\pi\right)^{-3}\sqrt{\mbox{Det}\{{\hat {\bf \Lambda}}\}}
  e^{-\frac 1 2 {\bf x}^T\hat{{\bf \Lambda}}{\bf x}}
\label{fo}
\end{eqnarray}
with
\begin{equation}
  {\hat {\bf \Lambda}}
  =
  \left(\begin{array}{cc}
    \mu  &0 \\
     0& \lambda
  \end{array}\right)
  \quad.
\end{equation}
Then the solution (\ref{sol}) of the selfconsistent Vlasov equation reads
\begin{eqnarray}
  &&f_{\delta}({\bf x},t)
  =
  \left(2\pi\right)^{-3}\sqrt{\mbox{Det}\{\hat{{\bf \Lambda}}\}}
  e^{-\frac 1 2 (\hat {\bf A} {\bf x}+{\bf C})^T{\hat {\bf \Lambda}}(\hat {\bf A} {\bf x}+{\bf C})}\nonumber\\
  &&=
  \left(2\pi\right)^{-3}\sqrt{\mbox{Det}\{\hat{{\bf \Lambda}}\}}
  e^{-\frac 1 2 ({\bf x}^T+{\bf C}^T(\hat {\bf A}^T)^{-1})\hat {\bf A}^T{\hat {\bf \Lambda}}\hat {\bf A}({\bf x}+\hat {\bf A}^{-1} {\bf C})}.\nonumber\\
\end{eqnarray}
We can now coarse grain the distribution with (\ref{eq:Husfold}) to obtain the result
\begin{eqnarray}
  f({\bf x},t)
  &=&
  {\left(2\pi\right)^{-3}   e^{-\frac 1 2 (\hat {\bf A} {\bf x}+{\bf C})^T(\hat {\bf \Lambda}^{-1}+\tilde { {\bf \Sigma}}^{-1})^{-1}(\hat {\bf A} {\bf x}+{\bf C})}
\over \sqrt{\mbox{Det}\{(\hat {\bf A}^T{\hat {\bf \Lambda}}\hat {\bf A})^{-1}+\hat { {\bf \Sigma}}\}}}
\nonumber\\
&&
\label{sola}
\end{eqnarray}
where
\beq
\tilde {\bf \Sigma}=\hat {\bf A} \hat {\bf \Sigma} \hat {\bf A}^{T}.\label{cos}
\eeq

We see that the coarse grained solution (\ref{sola}) can be represented at any time by the coarse grained initial distribution
\beq
f({\bf x},t)&=&\{f_{\delta}({\bf x},t) \star {\cal G}({\bf x})\}({\bf y})\nonumber\\
&=&\{f_0(\hat {\bf A} {\bf x} + {\bf C})\star {\cal G}({\bf x},t)\} ({\bf y})\nonumber\\
&=&\{f_0({\bf x})\star \tilde {\cal G}({\bf x},t)\} (\hat {\bf A} {\bf y}+{\bf C})\label{40}
\eeq
if we use the coarse graining (\ref{eq:Husfold}) but with the ( now time dependent) width
(\ref{cos}). It is immediately obvious that the last identity in (\ref{40}) is only valid if we use Gaussian initial distributions. Any other distributions will not allow this rearrangement. Consequently the folding and the dynamics are not interchangeable in general.

The entropy is calculated from (\ref{sola})
\begin{equation}
  S
  =
  3\log{(2\pi e)}
  +\frac 1 2 \log{{\rm Det} ((\hat {\bf A}^T{\hat {\bf \Lambda}}\hat {\bf A})^{-1}+\hat {{\bf \Sigma}})}
  \quad
\label{entra}
\end{equation}
with the explicit expression (\ref{entrb}).

\subsubsection{First coarse graining than solving}

Equation (\ref{vlas1}) is transformed into a partial differential equation of first order by Fourier transform ${\bf x}=(r,p)\rightarrow {\bf q} =(q,x)$. The solution can be obtained for the Fourier transformed distribution $\tilde f$
\beq
\tilde f({\bf q},t)=\tilde f_0((A^{-1})^T{\bf q}) \exp(\frac 1 2 {\bf q}^T \hat {\bf B} {\bf q} +{\bf D} {\bf q})
\eeq
with the matrix $\hat {\bf A}$ as defined in (\ref{coord1}) and
\beq
{\bf D}&=&i v a \int\limits^t d t' Q(t')\left ( \matrix{{\sin(\omega (t-t')) \over m\omega}\cr \cos(\omega (t-t'))} \right )
\nonumber\\
&&\nonumber\\
\hat {\bf B}&=&\frac 1 2({\sigma_p^2\over m\omega}-\sigma_r^2 m\omega)\left (\matrix{{1-\cos(2\omega t) \over m\omega}&\sin(2\omega t)\cr
\sin(2\omega t)&m\omega (1-\cos(2 \omega t))}\right ).\nonumber\\&&
\label{h1}
\eeq
The selfconsistency requirement (\ref{self}) leads to the same
solution as (\ref{se}). This shows that the coarse graining does not
affect the nonlinear feedback of the selfconsistent potential within
this model. In other more nontrivial models this needs not to be the case. 

If we now use as initial distribution once more a Gaussian (\ref{fo}) which reads in Fourier transform
\beq
\tilde f_0({\bf q})=\exp(-\frac 1 2 {\bf q}^T \hat {\bf \Lambda}^{-1} {\bf q})
\label{ftfo}
\eeq
we obtain the time dependent distribution after inverse Fourier transform
\beq
f({\bf x})={(2 \pi)^{-3}\over \sqrt{{\rm Det}\{\hat M\}}}\exp{\left
    [\!-\!\frac 1 2 ({\bf x} \!-\!i {\bf D})^T \hat {\bf M}^{-1}({\bf
      x} \!-\!i {\bf D})\right ]}
\label{solb}
\eeq
with
\beq
\hat {\bf M}=(\hat {\bf A}^T \hat {\bf \Lambda} \hat {\bf A})^{-1} -\hat {\bf B}.
\label{M}
\eeq



\begin{thebibliography}{10}

\bibitem{BCR92}
G.~F. Burgio, P. Chomaz, and J. Randrup, Phys. Rev. Lett. {\bf 69},  885
  (1992).

\bibitem{P89}
G. Peilert {\it et~al.}, Phys. Rev. C {\bf 39},  1402  (1989).

\bibitem{HM87}
B. ter Haar and R. Malfliet, Phys. Lett. B {\bf 172},  10  (1986).

\bibitem{BKG84}
G.~F. Bertsch, H. Kruse, and S. DasGupta, Phys. Rev. C {\bf 29},  673  (1984),
  errata: 33 (1986) 1107.

\bibitem{BD88}
G.~F. Bertsch and S.~D. Gupta, Phys. Rep. {\bf 160},  189  (1988).

\bibitem{D84}
P. Danielewicz, Ann. Phys. (NY) {\bf 152},  239  (1984).

\bibitem{GRSVR87}
C. Gregoire {\it et~al.}, Nucl. Phys. A {\bf 465},  317  (1987).

\bibitem{Ba92}
W. Bauer, Nucl. Phys. A {\bf 538},  83  (1992).

\bibitem{AB85}
J. Aichelin and G. Bertsch, Phys. Rev. C {\bf 31},  1730  (1985).

\bibitem{KWB92}
D. Klakow, G. Welke, and W. Bauer, Phys. Rev C {\bf 48},  1982  (1993).

\bibitem{MLSCN98}
K. Morawetz {\it et~al.}, Phys. Rev. Lett. {\bf 82},  3767  (1999).

\bibitem{KN84}
H.~S. K{\"o}hler and B.~S. Nilsson, Nucl. Phys. A {\bf 417},  541  (1984).

\bibitem{D84a}
P. Danielewics, Phys. Lett. B {\bf 146},  168  (1984).

\bibitem{K85}
H.~S. K{\"o}hler, Nucl. Phys. A {\bf 440},  165  (1984).

\bibitem{AK93}
M.~M. Abu-Samreh and H.~S. K{\"o}hler, Nucl. Phys. A {\bf 552},  101  (1993).

\bibitem{KB89}
H.~S. K{\"o}hler and W. Bauer, Phys. Rev. C {\bf 40},  1711  (1989).

\bibitem{N28}
L.~W. Nordheim, Proc. Roy. Soc. (A) {\bf 119},  689  (1928).

\bibitem{UU33}
E.~A. Uehling and G.~E. Uhlenbeck, Phys. Rev. {\bf 43},  552  (1933).

\bibitem{GBD87}
C. Gale, G. Bertsch, and S. DasGupta, Phys. Rev. C {\bf 35},  1666  (1987).

\bibitem{KJS85}
H. Kruse, B.~V. Jacak, and H. Stocker, Phys. Rev. Lett. {\bf 54},  289  (1985).

\bibitem{SG86}
H. St{\"o}cker and W. Greiner, Phys. Rep. {\bf 137},  277  (1986).

\bibitem{CLG87}
J. Cugnon, A. Lejeune, and P. Grang\'e, Phys. Rev. C {\bf 35},  861  (1987).

\bibitem{SLM96}
V. {\v S}pi{\v c}ka, P. Lipavsk{\'y}, and K. Morawetz, Phys. Lett. A {\bf 240},
   160  (1998).

\bibitem{MLSK98}
K. Morawetz, P. Lipavsk{\'y}, V. {\v S}pi{\v c}ka, and N.~H. Kwong, Phys. Rev.
  C {\bf 59},  3052  (1999).

\bibitem{MLNCCT01}
K. Morawetz {\it et~al.}, Phys. Rev. C {\bf 63},  034619  (2001).

\bibitem{BKN76}
P. Bonche, S. Koonin, and J. Negele, Phys. Rev. C {\bf 13},  1226  (1976).

\bibitem{W82}
C.~Y. Wong, Phys. Rev. C {\bf 25},  1460  (1982).

\bibitem{BBS84}
A. Bonasera, G.~F. Bertsch, and E.~N. El-Sayed, Phys. Lett. B {\bf 141},  9
  (1984).

\bibitem{GG96}
M. Gro{\ss{}} and C. Guet, Phys. Rev. A {\bf 54},  R2547  (1996).

\bibitem{PG99}
L. Plagne and C. Guet, Phys. Rev. A {\bf 59},  4461  (1999).

\bibitem{CRSU00}
F. Calvayrac, P.~G. Reinhard, E. Suraud, and C.~A. Ullrich, Phys. Rep {\bf
  337},  493  (2000).

\bibitem{JB96}
C. Jarzynski and G.~F. Bertsch, Phys. Rev. C {\bf 53},  1028  (1996).

\bibitem{K75}
Y.~L. Klimontovich, {\em Kinetic theory of nonideal gases and nonideal plasmas}
  (Academic Press, New York, 1975).

\bibitem{LSM97}
P. Lipavsk{\'y}, K. Morawetz, and V. {\v S}pi{\v c}ka, {\em Kinetic equation
  for strongly interacting dense Fermi systems} (Annales de Physique, Paris,
  2001), No.~{26, 1}, {K}. Morawetz, Habilitation University Rostock 1998.

\bibitem{G5}
J.~W. Gibbs,  in {\em Elementare Grundlagen der statistischen Mechanik} (J. A.
  Barth, Leipzig, 1905).

\bibitem{EE7}
P. Ehrenfest and T. Ehrenfest, {\em Enzyklop{\"a}die der Mathematischen
  Naturwissenschaften} (1907-1914), No.~4, p.\ Art. 32.

\bibitem{S76}
D.~N. Zubarev, {\em Statistische Thermodynamik des Nichtgleichgewichts}
  (Akademie Verlag, Berlin, 1976).

\bibitem{RS95a}
P.~G. Reinhard and E. Suraud, Ann. of Phys. {\bf 239},  193  (1995).

\bibitem{RS95b}
P.~G. Reinhard and E. Suraud, Ann. of Phys. {\bf 239},  216  (1995).

\bibitem{LSR95}
P. L'Eplattenier, E. Suraud, and P.~G. Reinhard, Ann. of Phys. {\bf 244},  426
  (1995).

\bibitem{T89}
K. Takahashi, Prog. of Theor. Phys. Suppl. {\bf 98},  109  (1989).

\bibitem{Mr97}
K. Morawetz, Phys. Rev. C {\bf 55},  R 1015  (1997).

\bibitem{MVFLS98}
K. Morawetz {\it et~al.}, Phys. Rev. C. {\bf 60},  54601  (1999).

\bibitem{M99}
K. Morawetz, Phys. Rev. E. {\bf 61},  2555  (2000).

\bibitem{MFW00}
K. Morawetz, U. Fuhrmann, and R. Walke,  in {\em Isospin Effects in Nuclei},
  edited by B.~A. Lie and U. Schroeder (World Scientific, Singapore, 2000),
  correct version: nucl-th/0001032.

\bibitem{OWW84}
R.~F. O'Connell, L. Wang, and H.~A. Williams, Phys. Rev. A {\bf 30},  2187
  (1984).

\end{thebibliography}

\end{document}